\documentclass[journal]{IEEEtran}
\ifCLASSINFOpdf
\else
\fi
\usepackage{cite}
\usepackage{amssymb}
\usepackage{multirow}
\usepackage{graphicx}
\usepackage{rotating}
\usepackage{xcolor}
\usepackage{fancyhdr}
\usepackage{amsmath}

\usepackage{algorithmic}
\usepackage{array}
\ifCLASSOPTIONcompsoc
\else
\usepackage[caption=false,font=footnotesize]{subfig}
\fi
\usepackage{fixltx2e}
\usepackage{stfloats}
\let\MYoriglatexcaption\caption
\renewcommand{\caption}[2][\relax]{\MYoriglatexcaption[#2]{#2}}
\usepackage{url}
\begin{document}

\title{Joint Design of Self-Tuning UHF RFID Antenna and Microfluidic Channel for Liquid Sensing}
\author{Giulio~Maria~Bianco,~\IEEEmembership{Member,~IEEE,}
    and~Gaetano~Marrocco,~\IEEEmembership{Senior~Memeber,~IEEE}
\thanks{Paper supported by Project ECS 0000024 Rome Technopole, CUP B83C22002820006, NRP Mission 4 Component 2 Investment 1.5, Funded by the European Union – NextGenerationEU. Rome Technopole Projects: ”Next-gen point of cares: chemical-physical sensors with wireless interface for health monitoring in domestic settings” (Spoke 1; Flagship Project 7) and ”Eco-friendly Electronic Labels for Food and Plastic Waste” (Spoke 2; Flagship Project 3).}
\thanks{G. M. Bianco is with the Department of Civil Engineering and Computer Science Engineering, the University of Rome Tor Vergata, Rome, Italy; e-mail: giulio.maria.bianco@uniroma2.it.}
\thanks{G. Marrocco is with the Department of Civil Engineering and Computer Science Engineering, the University of Rome Tor Vergata, Rome, Italy.}
\thanks{Manuscript received Month XX, 20XX; revised Month XX, 20XX.}}


\maketitle

\begin{abstract}
Microfluidic has been an enabling technology for over a decade, particularly in the field of medical and wearable devices, allowing for the manipulation of small amounts of fluid in confined spaces. Micro-channels can also be used for wireless sensing thanks to the variations in antenna properties when the fluid flows near it. However, up to now, microfluidic channels and sensing antennas have always been designed separately; instead, since the liquid flow and the antenna geometry both contribute to the overall performance, they should be considered simultaneously when optimizing the antenna-microfluidic system. In this paper, the joint design of the antenna and microfluidic channels is investigated for liquid quantification. Self-tuning RFID microchips are exploited to minimize communication degradation due to the increase of lossy liquid amount over the sensing antenna while digitalizing the impedance mismatch itself. To experimentally corroborate the joint design technique, two different geometries are obtained and prototyped starting from a given antenna-microfluidic layout by setting different goals for an optimization function. The two flexible RFID prototypes returned performance in agreement with the simulated ones, achieving a maximum sensitivity of about $20$~units of the digital metric per milligram increase of water.
\end{abstract}


\section{Introduction}

\IEEEPARstart{M}{icrofluidic} has been standing as one of the most impactful enabler technologies for more than a decade, mostly thanks to the advancements in medical and wearable devices~\cite{Chen19}. Microfluidic channels allow for manipulating a small amount of fluid in a confined space, the scale typically of a few to
hundreds of microliters, and they can be integrated into small devices for performing even complex in-loco sensing~\cite{Li20}, usually through electrochemical sensors~\cite{Khashayar22}. The adoption of microfluidic naturally involved wireless devices in three main research paths: \textit{i})~the development of antennas made of liquid metal, usually exploiting microfluidic to achieve spatial reconfigurability~\cite{Ren23, Mohamadzade20, Paracha20}, and use of microfluidic \textit{ii})~to acquire sensory data, eventually by chemical reactions~\cite{Fiore23, Ria23, Riente23, Chen19}, and \textit{iii})~to produce variations in the antenna's properties for sensing scopes~\cite{Zhu20, Seo16, Mariotti15}.\par

Concerning liquid and humidity sensing, remote monitoring based on variations in antennas' properties has been extensively investigated in recent years as well, especially by means of RFID (radiofrequency identification) technology. Several chipped~\cite{Tajin21, Chen20} 
and chipless~\cite{Xue23, Marchi23} solutions showed promising sensitivity but suffered from a significant reduction in antenna gain and, hence, in the reading distances. Thus, a peculiar kind of self-tuning RFID ICs (integrated circuits) was proposed to complete sensing without sacrificing the reading distances excessively~\cite{Bianco20Near}. Indeed, such microchips can compensate for limited impedance mismatches and return a digital metric proportional to the retuning effort. 

Table~\ref{tab:biblio} summarizes the main, recent papers on the joint use of antennas and microfluidic including the maximum volume of liquid considered\footnote{When the liquid volume was not reported directly in the reference, it was evaluated from the geometrical parameters of the microchannel.}. Although liquid sensing is the main application of the research topic, the microfluidic subsystem has always been considered divided from the microwave one so that one of the two subsystems is designed only after having fixed the other. Therefore, the joint design of the antenna and the microfluidic channels to achieve optimal performance has never been investigated.\par

\begin{figure}[tp]
\centering
\includegraphics[width=6 cm]{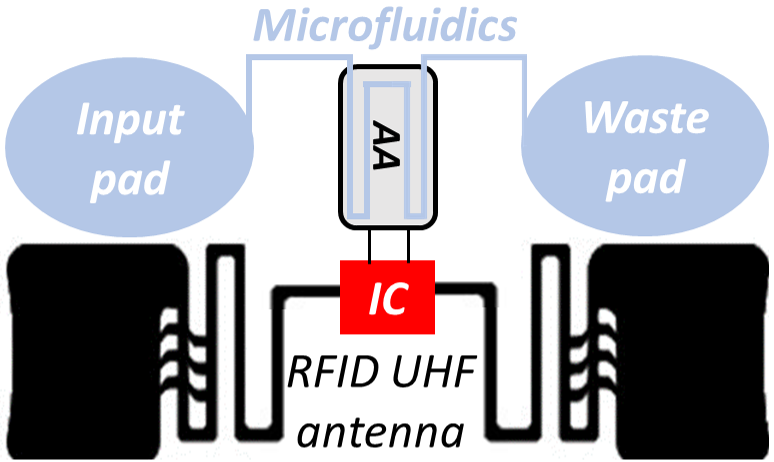}
\caption{ Concept of a UHF RFID sensor-antenna with microfluidic for liquid quantification. The sensitive area of the system is constituted by the superposition of the microfluidic channel with the AA.}
\label{fig:concept}
\end{figure}

\begin{table*}[!t]
\caption{Comparison between this research and recent works on microfluidic-antenna systems.}\label{tab:biblio}
\begin{scriptsize}
    \centering
    \begin{tabular}{l||l|l|l|l|l|l|l}
        \textbf{Reference} & & \textbf{Microfluidic role} & \textbf{Working} & \textbf{Maximum} & \textbf{Digitalized} & \textbf{Joint} & \textbf{} \\
        \textbf{(Year)} & \textbf{Scope} & \textbf{(liquid quantity $\mu$L)} & \textbf{frequency} & \textbf{size} & \textbf{sensing} & \textbf{design} & \textbf{Description} \\
    \hline
    \hline
        \textbf{This work} & Remote sensing & Govern liquid& & $50$~mm~$\times$ & & & Joint antenna-microfluidic \\
        \textbf{(2024)} & (liquid quantity) & flow ($1050$)& $925$~MHz & $20$~mm & Yes & Yes & design to govern performance\\
    \hline
         \cite{Amiri22} & Remote sensing & Govern & $3.12$~GHz to & $140$~mm~$\times$  & & & Metamaterial sensor for remote \\
         (2022) & (water salinity) & liquid flow ($6$)& $3.59$~GHz & $160$~mm & No & No & sensing of water salinity levels \\
    \hline
        \cite{Zhu20}  & Remote sensing & Govern & $2.9$~GHz and & $14$~mm~$\times$  & & & Harmonic-based sensor with \\
        (2020)& (liquid quantity) & liquid flow ($630$) & $5.7$~GHz & $6$~mm  & No & No & a micromachined liquid channel \\
    \hline
        \cite{Silavwe16} & Remote sensing & Govern & $8.8$~GHz to & $5$~mm~$\times$ & & & Microwave-integrated sensor for \\
        (2016)& (liquid characterization) & liquid flow ($7$) & $9.0$~GHz & $12$~mm & No & No & liquid sample characterization \\
    \hline
        \cite{Seo16}  & Remote sensing & Govern & $4.2$~GHz to & $13$~mm~$\times$ & & & Substrate-integrated waveguide \\
        (2016)& (ethanol quantity) & liquid flow ($1$) & $4.6$~GHz & $19$~mm & No & No & for chemical sensing \\
    \hline
        \cite{Yuan16}  & Remote & Govern & $840$~MHz to & $95$~mm~$\times$ & & & Biosensor with microfluidic \\
        (2016)& biosensing & liquid flow ($120$) & $960$~MHz & $8$~mm & No & No & and self-assembling antennas \\
    \hline
        \cite{Pradhan23}  & Reconfigurable & Change & $3.7$~GHz and & $40$~mm~$\times$ & & & Self-diplexing antenna based on \\
        (2023)& antenna & frequency ($75$) & $5.9$~GHz & $40$~mm & - & No & air- and liquid-filled pockets \\
    \hline
        \cite{Sanusi22}  & Reconfigurable & Reconfiguring meta- & $8$~GHz to & $200$~mm~$\times$ & & & Reconfigurable metasurface \\
        (2022)& metamaterial & metamaterial ($24375$) & $13$~GHz & $200$~mm & - & No & based on liquid metal injection \\
    \hline
         \cite{Goode21} & Reconfigurable & Reconfiguring & $32$~GHz to & $45$~mm~$\times$ & &  & Reconfigurable antenna \\
        (2021)& antenna & antenna ($703$) & $35$~GHz & $20$~mm & - & No & with fluidically-loaded lens \\
    \hline
        \cite{Liu21}  & Reconfigurable & Reconfiguring & $8$~GHz to & $40$~mm~$\times$ & & & Liquid metal alloy and micro-\\
        (2021)& array & metasurface ($216$) & $32$~GHz & $40$~mm & - & No & fluidics for a metasurface \\
    \hline
        \cite{Gonzalez18}  & & Feed network & & $3$~mm~$\times$~$2$~mm & & & Microfluidically-switched feed \\
        (2018)& Beam steering & switching ($395$) & $30$~GHz & (single element) & - & No & for beam-steering \\
    \end{tabular}
\end{scriptsize}
\end{table*}

In this work, we propose a global design technique wherein the geometrical parameters of the channel and those of the antenna are simultaneously optimized by numerical simulations to quantify a liquid volume. Indeed, liquid sensing is based on the flow of the liquid onto the antenna so that the electromagnetic variations will be a function of the geometrical parameters of both the antenna and the microfluidic shape. The gain degradation expected by the liquid’s absorption of electromagnetic waves and the antenna detuning is minimized thanks to a self-tuning UHF (ultra-high frequency) RFID IC. Moreover, the flow of the liquid is governed by a microfluidic channel superimposed on the AA (antenna impedance adapter) to dominate the detuning of the antenna~(Fig.~\ref{fig:concept}). This sensing mechanism is digitalized and returned by the IC, avoiding known issues in similar analog sensing approaches~\cite{Occhiuzzi16}. Accordingly, the joint design of the sensing antenna and the microfluidic channel has to be conceived and validated to optimize the overall performance of the system that is simultaneously function of both geometries. \par

The paper's organization follows. Section~\ref{sec:symbols} lists the most relevant symbols used in this paper; afterwards, Section~\ref{sec:background} recalls the known scientifical background on the self-tuning RFID ICs and on microfluidics. The joint design problem is formulated and then managed through a goal function in Section~\ref{sec:jointDesign}. Section~\ref{sec:corroboration} presents a design example by numerical simulations, whereas the relative experimental corroboration of the proposed design technique follows in Section~\ref{sec:prototypation}.

\section{Symbols}\label{sec:symbols}

For the reader’s convenience, the most relevant symbols used in the remainder of the paper are listed in this Section in order of appearance.

\begin{table}[h!]
    \begin{tabular}{ll}
        $C_s$ & Self-tuning capacitance step \\
        $C_0$ & Self-tuning baseline capacitance \\
        $C_{IC}\left(s\right)$ & Self-tuning overall capacitance \\
        $s$ & Sensor code (SC)\\
        $\omega$ & Frequency pulse \\
        $B_A$ & Antenna susceptance \\
        $s_{min}$ & Lowest value of the SC linear range\\
        $s_{max}$ & Highest value of the SC linear range\\
        $\Psi$ & Measurand\\
        $\Delta s$ & Differential sensor code\\
        $s_0$ & Reference value of the differential sensor code\\
        $\boldsymbol{v}$ & Vector of variable geometrical parameters for the joint design\\
        $a_i$ & $i$-th variable geometrical parameter of the antenna\\
        $c_i$ & $i$-th variable geometrical parameter of the microfluidic channel\\
        $\xi$ & Quantity of liquid in the microfluidic\\
        $\xi_0$ & Unloaded condition of the microfluidic, viz., in the absence \\ 
        & of liquid (initial state)\\
        \end{tabular}
    \end{table}
\begin{table}[h!]
    \begin{tabular}{ll}
        $\xi_F$ & Loaded condition of the microfluidic, viz., microfluidic filled \\ 
        & with liquid completely (final state)\\
        $F\left[\boldsymbol{v}\right]$ & Fitness function for the joint design, sum of $f_i\left[\boldsymbol{v}\right]$ functions\\
        $w_i$ & $i$-th weight of the fitness function $f_i\left[\boldsymbol{v}\right]$\\
        $G_{\tau}\left(\xi\right)$ & Realized radiation gain in presence of liquid ($\xi$)\\
        $S\left[\boldsymbol{v}\right]$ & Sensitivity of the antenna-microfluidic system\\
        $c_1$ & Thickness of the microfluidic\\
        $\varepsilon$ & Dielectric constant\\
        $\tan\delta$ & Loss tangent\\
        $a_1$ & Side of the $\Gamma$-match transformer parallel to the IC\\
        $a_2$ & Side of the $\Gamma$-match transformer perpendicular to the IC\\
        $c_2$ & Width of the microfluidic channel\\
        $a_3$ & Width of the antenna's conductor trace\\
        $G_{IC}$ & Conductance of the integrated circuit\\
        $c_3$ & Side of the microfluidic parallel to the IC\\
        $c_4$ & Step of the microfluidic serpentine\\
        $\rho_l$ & Density of the fluid\\
    \end{tabular}
\end{table}

\section{Background}\label{sec:background}

\subsection{Self-tuning RFID ICs}\label{sec:backgroundSelftuning}

Self-tuning microchips modify their internal capacitance to maximize the power delivered by the harvesting antenna to the IC itself~\cite{Bianco20Near, Caccami18}. By assuming a constant incremental step $C_s$ and denoting the baseline capacitance $C_0$, they are modelled as a resistance in parallel to the variable capacitance $C_{IC}\left(s\right)$ where $s$ is named \emph{sensor code}~(SC), which is an adimensional digital unit. The self-tuning equation determines the actual value of $C_{IC}$ based on the frequency pulse $\omega$ and the antenna susceptance $B_A$
\begin{equation}\label{eq:STequation}
    \left|\omega C_{IC}\left(s\right)+B_A\right|=0.
\end{equation}
Equation~(\ref{eq:STequation}) is fully valid for a microchip-dependent SC range $s_{min}\leq s\leq s_{max}$. If the antenna susceptance is a function of a measurand $\Psi$, the SC can be exploited for sensing through~(\ref{eq:STequation})~\cite{Naccarata22}:
\begin{equation}\label{eq:SCevaluation}
    s\left(\Psi\right)=\textnormal{nint}\left\{ -\frac{1}{C_s}\left[C_{IC}\left(s_{min}\right)+\frac{B_A\left(\Psi\right)}{\omega}\right] \right\}.
\end{equation}
Measurand-independent baselines can be removed by resorting to the \emph{Differential sensor code}~($\Delta s$) w.r.t. a reference value $s_0$~\cite{Naccarata22}.

\subsection{Microfluidic Channels}\label{sec:microfluidic}

\begin{figure}[tp]
\centering
\includegraphics[width=5 cm]{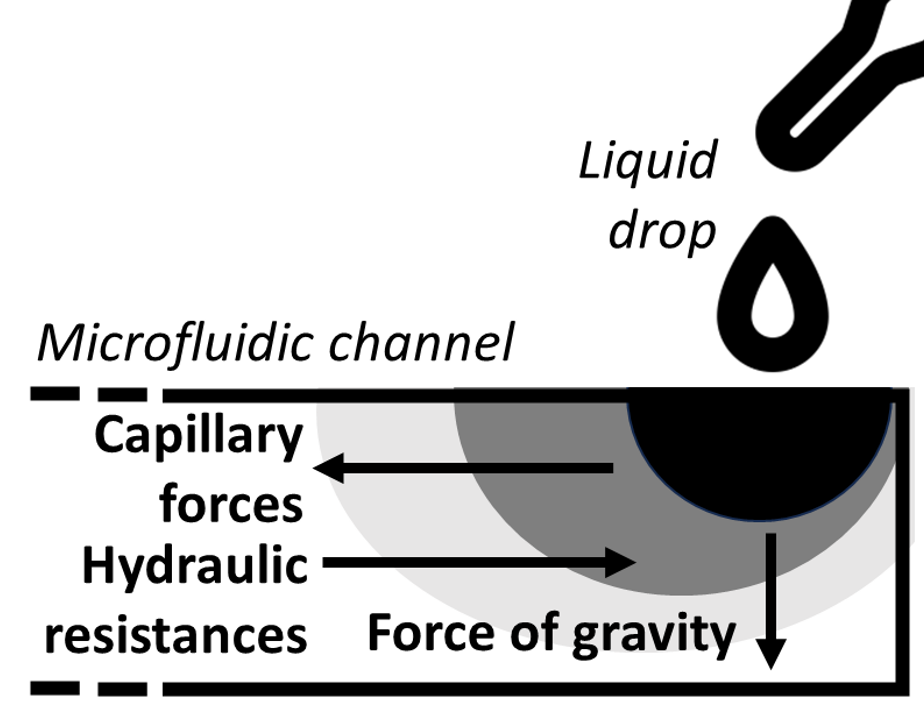}
\caption{Model of the fluid diffusion in paper-based microfluidic.}
\label{fig:capillaryFlow}
\end{figure}

Microfluidic is a topic of research in the manipulation of microliters, encompassing theoretical physics, numerical methods, fabrication, and deployment~\cite{Mitra11}. Hereafter, the particular case of \textit{capillary pumps} is exploited. Capillary pumps are microchannels placed horizontally wherein the liquid propagates by capillarity, and since there is no opposition from gravity, the (ideal) capillary flow continues until the channel is completely traversed thanks to the capillary forces~\cite{Bruus06}. In real microfluidic, hydraulic resistances caused by the interaction with the channel's walls oppose the capillary rise, and the liquid will also flow downward due to gravity if it is not opposed by any solid surface~\cite{Bruus06} (Fig.~\ref{fig:capillaryFlow}). Specifically, in this work, we consider Whatman CF$4$ paper which tends to have variable flow rates due to the structure of the paper fibers so that fixed liquid quantities might flow to different positions each time.

\section{Joint Antenna-Microfluidic Design} \label{sec:jointDesign}

In this section, the joint design technique is introduced and addressed through a goal function. 

\subsection{Problem Formulation of the Joint Design}

The goal of the joint design of the antenna and the microfluidic is, hence, to quantify a small volume of water having the order of magnitude of a few microliters while ensuring a stable reading range of the RFID sensor. To achieve this objective, a self-tuning RFID IC and microfluidic channels can be employed. Then, the problem is formulated as the optimal design of the antenna-microfluidic pair to \emph{i})~minimize the performance degradation due to the increasing, lossy liquid, and \emph{ii})~maximize the sensitivity of the antenna's response to the liquid quantity.\par

Let us assume that the antenna and microfluidic channels, including the materials and placement of the RFID sensor, are parametrized by a vector of variable geometrical parameters~$\boldsymbol{v}$
\begin{equation}\label{eq:vVector}
    \boldsymbol{v}=\left\{a_1,...,a_{N},c_1,...,c_{M}\right\}
\end{equation}
where $\left\{a_1,...,a_N\right\}$ are the antenna's parameters, whereas $\left\{c_1,...,c_M\right\}$ are the microfluidic channels' ones. The hyperspace $\left\{a_{1,min}\leq a_1 \leq a_{1,max},...,c_{M,min}\leq c_M \leq c_{M,max}\right\}$ to be explored for finding the best solution is naturally delimited by size and manufacturing constraints.\par

Therefore, the liquid quantity $\xi$ increases with time between the starting and final instances of the phenomenon to be monitored. Hence,
\begin{equation}\label{eq:xiTime}
    \xi_0 < \xi < \xi_F
\end{equation} 
being $\xi_0$ and $\xi_F$ the conditions of channel fully free and filled with liquid, respectively. When the liquid increases, the susceptance's variations must be monotonic to perform sensing by self-tuning correctly
\begin{equation} \label{eq:workingPoint}
\begin{split}
B_A\left(\xi_0\right)< B_A\left(\xi\right) < B_A\left(\xi_F\right)  \veebar \\
B_A\left(\xi_0\right)> B_A\left(\xi\right)> B_A\left(\xi_F\right)
\end{split}
\end{equation}
where $\veebar$ is the XOR (exclusive or) boolean operator. This must be imposed when selecting the general antenna-microfluidic geometry to be optimized in the given parameters' space.

\subsection{Boundary Conditions and Goal Function}

The optimal layout can be determined by considering the initial ($\xi_0$) and final ($\xi_F$) states of the sensor. Then, the optimization problem can be formalized as the maximization of a three-term \emph{fitness function (F)} as follows
\begin{equation} \label{eq:fitnessFunction}
F\left[\boldsymbol{v}\right]=\left \{ \begin{array}{ll}
\left(\sum_{i=1}^3w_i\right)^{-1}\sum_{i=1}^3w_if_i\left[\boldsymbol{v}\right] \\
\textnormal{if} \quad \left(f_1 \land f_2 \land f_3\right) \neq 0 \\
0 \qquad \textnormal{elsewhere}
\end{array}
\right.
\end{equation}
where $\land$ is the AND boolean operator, and
\begin{equation} \label{eq:sub1}
f_1\left[\boldsymbol{v}\right]=\left \{ \begin{array}{ll}
\frac{s\left[\boldsymbol{v}\right]\left(\xi_0\right)}{s_{max}} \qquad \textnormal{if} \quad s\left[\boldsymbol{v}\right]\left(\xi_0\right) \leq s_{max} \\
0 \qquad \textnormal{elsewhere}
\end{array}
\right.
\end{equation}
\begin{equation} \label{eq:sub2}
f_2\left[\boldsymbol{v}\right]=\left \{ \begin{array}{ll}
\frac{G_{\tau}\left[\boldsymbol{v}\right]\left(\xi_0\right)+G_{\tau}\left[\boldsymbol{v}\right]\left(\xi_F\right)}{2G_{0}} \\
\textnormal{if} \quad \textnormal{min}\left(G_{\tau}\left[\boldsymbol{v}\right]\left(\xi_0\right),G_{\tau}\left[\boldsymbol{v}\right]\left(\xi_F\right)\right)\geq G_{min}\\
0 \qquad \textnormal{elsewhere}
\end{array}
\right.
\end{equation}
\begin{equation} \label{eq:sub3}
f_3\left[\boldsymbol{v}\right]=\left \{ \begin{array}{ll}
\frac{S\left[\boldsymbol{v}\right]}{S_0} \qquad \textnormal{if} \quad S\left[\boldsymbol{v}\right] \geq S_{min}\\
0 \qquad \textnormal{elsewhere}
\end{array}
\right. .
\end{equation}
The notation $s\left[\boldsymbol{v}\right]\left(\xi\right)$ indicates the value of $s$ when the liquid in the channel is $\xi$ while considering the $\boldsymbol{v}$ vector of parameters. $\left\{w_1,w_2,w_3\right\}$ are a set of weights, $S$ is the sensitivity, which is measured in SC units over milligrams of liquid, $G_{\tau}\left(\xi\right)$ is the realized gain in the presence of liquid's volume $\xi$, $\left\{G_{min}, S_{min}\right\}$ are the minimum acceptable performance value that automatically discards solutions if they are not met, and $S_0$ and $G_0$ are two normalization values such that each addend of $F\left[\boldsymbol{v}\right]$ is comprised between $0$~and~$1$. The sensitivity to the liquid quantity is defined as a function of the overall weight of water in the sensitive area in loaded conditions $\xi_F$
\begin{equation}\label{eq:sensitivity}
S\left[\boldsymbol{v}\right]=\frac{\left|s\left[\boldsymbol{v}\right]\left(\xi_F\right)-s\left[\boldsymbol{v}\right]\left(\xi_0\right)\right|}{\xi_F\left[\boldsymbol{v}\right]}.
\end{equation}

The sub-term~(\ref{eq:sub1}) imposes that the whole flow of interest is captured by the antenna in the case of monotonically-decreasing susceptance\footnote{The case of monotonically-increasing susceptance can be derived straightforwardly.} so that the dynamic range of the sensor is maximized, whereas the sub-term~(\ref{eq:sub2}) imposes a check on the reading distance of the RFID tag while the liquid increases. The sensitivity to the liquid quantity, which depends on both the antenna response and microfluidic geometry, is accounted for by (\ref{eq:sub3}) and (\ref{eq:sensitivity}). The terms can be weighted according to the application's needs. It is worth highlighting that both the geometrical parameters of the antenna and microfluidic affect all three fitness terms (\ref{eq:sub1})-(\ref{eq:sub3}) due to the close interrelationship between the microwave and the microfludic subsystems.

\section{Design Example}\label{sec:corroboration}

The joint design is demonstrated here with two antenna-microfluidic systems by starting from the same general geometry and applying different weightings in (\ref{eq:fitnessFunction}).

\subsection{System Layout and Parameters' Space}

\begin{figure}[!t]
\centering
\includegraphics[width=7.5cm]{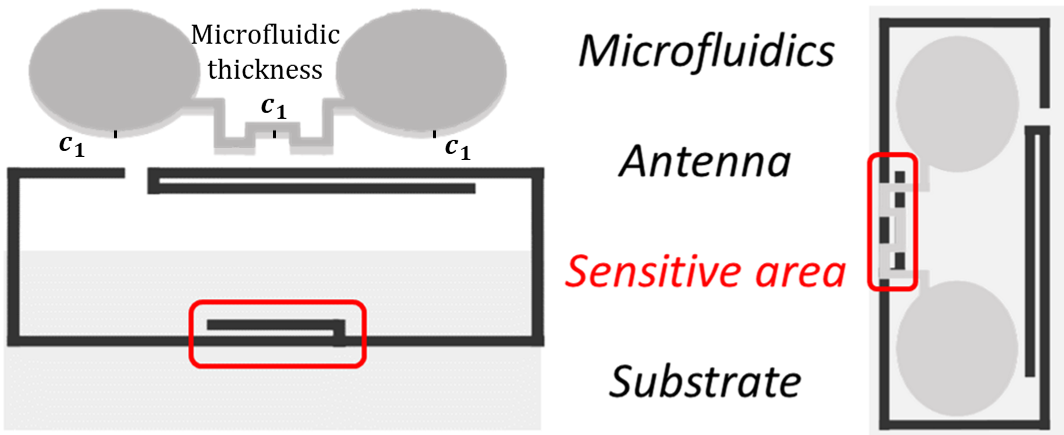}\\
\qquad \qquad (a) \qquad \qquad \quad \qquad \qquad \qquad \quad (b) \\
\includegraphics[height=5cm]{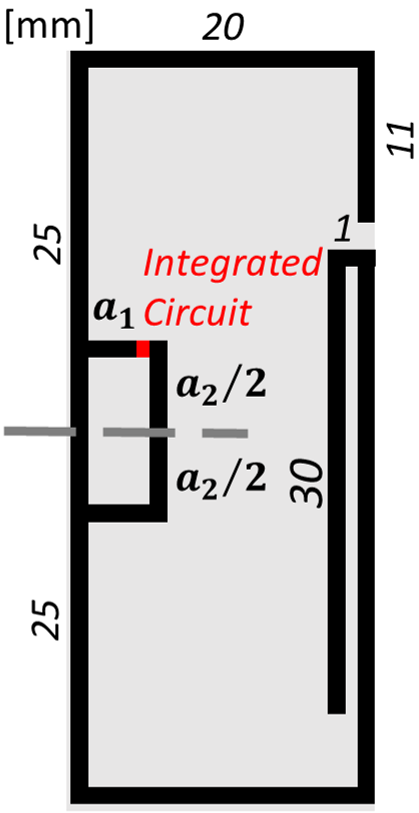}\quad\includegraphics[height=4.7cm]{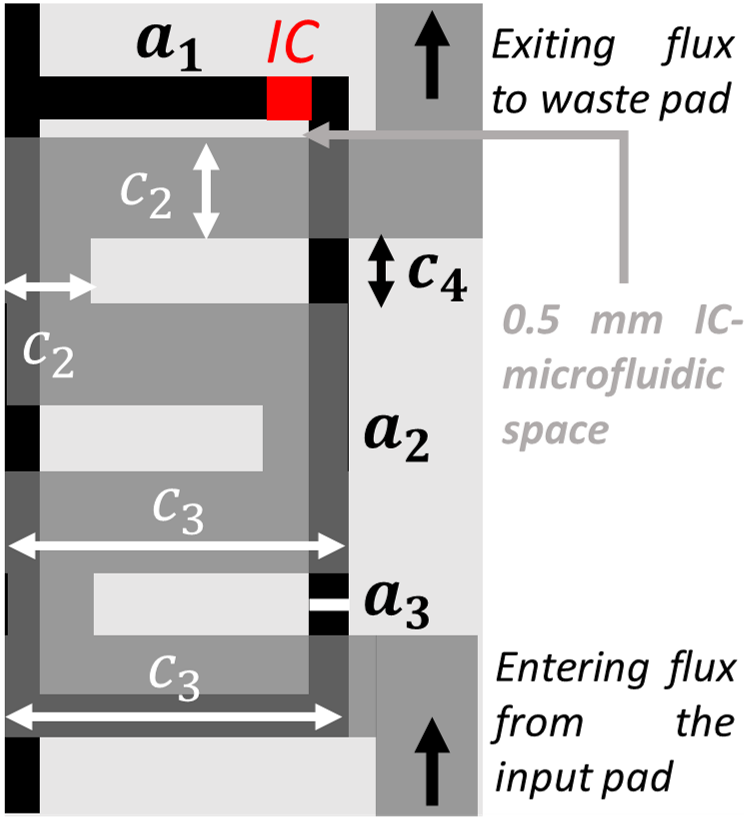}\\
\quad (c) \qquad \qquad \qquad \quad (d) \qquad \qquad \qquad \qquad
\caption{System layout composed by an RFID sensor-antenna and microfluidic for liquid quantification. The sensitive area of the system is highlighted in red. (a)~Exploded view with highlighted AA and $c_1$ parameter, and (b)~above view with the highlighted sensitive area. Parametrization of the antenna-microfluidic system. (c) Geometrical parameters of the RFID antenna and the pair to be optimized $\left\{a_1,a_2\right\}$. (d) Zoomed-in view of the system's sensitive area and geometrical parameters of the microfluidic channels to be optimized $\left\{c_2,c_3,c_4\right\}$. $\left\{a_3,c_1\right\}$ are fixed.}
\label{fig:parametriz}
\end{figure}

\begin{table}[t]
\caption{Weights and parameters' values.}
\centering
\begin{tabular}{l||r|r|r|r|r|r|r|r}\label{tab:weightsGeom}
        \textbf{Geometry} & $\mathbf{w_1}$ & $\mathbf{w_2}$ & $\mathbf{w_3}$ & $\mathbf{a_1}$ & $\mathbf{a_2}$ & $\mathbf{c_2}$ & $\mathbf{c_3}$ & $\mathbf{c_4}$ \\
        \hline
        \hline
        Sensitivity-opt. & $1$ & $1$ & $1$ & $0$ & $10.5$ & $2$ & $3$ & $0.3$\\
        Gain-optimized & $0$ & $5$ & $1$ & $6$ & $5$ & $1.5$ & $9$ & $0$ \\
\end{tabular}
\end{table}

The starting geometry and parameters' space were selected by considering a flexible RFID sensor for monitoring a small water content in a compact $50$~mm~$\times$~$20$~mm surface. The microfluidic has thickness $c_1=1$~mm and comprises two identical ellipses (that are neglected in the electromagnetic simulations) as collecting and waste pads' sections and by a serpentine passing on the antenna's AA (Fig.~\ref{fig:parametriz}(a,b)). In this example, the minimum volume of liquid to be monitored is $10$~mm$^3$ which is a microvolume suitable for initiating monitoring the sweating rate during domestic fitness~\cite{Bianco23PoC} depending on the body area and exercise routine~\cite{Baker18}. Ovoidal input and waste pads are used as examples since the design of the pads to carry the liquid of interest does not have any effect on the EM (electromagnetic) response of the antenna-microfluidic system and it is, hence, outside the scope of this work.\par

The antenna geometry is a partially folded $\Gamma$-match dipole on a thin layer of Kapton ($\varepsilon=3.5$, tan$\delta=0.0026$, thickness $0.125$~mm). The antenna is made of a copper trace (thickness $35$~$\mu$m) and can host the whole microfluidic within its area, including the collecting and waste pads. The dipole length is such to match the conductance of the chosen self-tuning IC (Magnus S3 by Axzon, conductance $G_{IC}=0.0482$~mS, capacitance ranging between $1.9$~pF and $2.9$~pF).\par

The parameters' space is such that the susceptance of the antenna moves around the linear SC range at the working frequency of $925$~MHz, which is an unlicensed UHF frequency commonly used for RFID applications in the United States of America. The width of the microfluidic paper trace is also bound to be larger than $0.5$~mm because of manufacturing constraints. The parameterization of the whole antenna-microfluidic geometry is summarized in Fig.~\ref{fig:parametriz}(c,d). A $1$-mm gap hosts the IC, and the free parameters to be optimized are the form factors of the $\Gamma$-match transformer ($\left\{a_1,a_2\right\}$) and the microfluidic trace width ($c_2$), whereas the copper trace width is fixed ($a_3=1$~mm) and the other variable parameters of the microfluidic are functions of the free parameters as follows
\begin{equation}\label{eq:alphaFormula}
c_3=a_1+2a_3+1,
\end{equation}
\begin{equation}\label{eq:gammaFormula}
c_4=\max\left(0,\frac{a_2-a_3-0.5-4c_2}{3}\right).
\end{equation}
For $c_4=0$, the serpentine degenerates in a straight channel. Consequently, given $\rho_l$ density of the fluid (for water, considered hereafter, $\rho_l=1$~mg/mm$^3$), the denominator in (\ref{eq:sensitivity}) is
\begin{equation} \label{eq:liquidWeight}
\begin{split}
\xi_F\left[\boldsymbol{v}\right]=\min\{\rho_lc_1c_3\left(a_2-a_3-0.5\right), \\
\rho_lc_1c_2\left(4c_3+3c_4\right)\}.
\end{split}
\end{equation}
From (\ref{eq:liquidWeight}), a suitable parameters' space to monitor at least $10$~mm\textsuperscript{3} is given by $0$~mm~$\leq a_1\leq8$~mm, $5$~mm~$\leq a_2\leq15$~mm, and $1$~mm~$\leq c_2\leq2.5$~mm.

\subsection{EM Modeling of the Microfluidic}\label{sec:emmodeling}

Hereafter, microfluidic is accounted for in electromagnetic models by an increasing quantity of the fluid. The channel material and fluid dynamics are assumed to be negligible; namely, the channel material does not significantly affect the antenna's susceptance, the transient regime of the liquid flow does not interfere with the self-tuning mechanism of the IC, and the liquid front propagates only along the direction of the microfluidics. These simplifications are valid for a wide range of practical cases, like the use of paper-based channels~\cite{Li23} for monitoring biological processes~\cite{Torii95}. \par

In the EM numerical simulations, therefore, the water ($\varepsilon=78$, $\sigma=1.78$~S/m) fills the channel's volume with a sharp liquid front. To perform the simulation-measurement comparison of the filling percentage of the channel despite the Whatman paper properties~(Sec~\ref{sec:microfluidic}), the weight of the antenna-microfluidic system is used. The weight corresponding to the empty channel was observed by adding water to the input pad until the sensitive channel was reached; then, the weight corresponding to the $100\%$-filled channel in the EM simulations was experimentally observed by adding more liquid until a stationary value of the SC was achieved, viz., the lowest SC level the system could return. Hence, the comparison between the simulated (through liquid front) and measured (by weighting) filling percentages could be drawn.\par

\subsection{Numerical Optimization}

\begin{table*}[t]
\caption{Simulated antenna performance. The normalization values are $G_0=2.50$~$\textnormal{dBi}$ and $S_0=19.4$~$\textnormal{mg}^{-1}$.}\label{tab:simPerf}
    \centering
    \begin{tabular}{l||r|r|r|r|r|r|r|r|r|r}
        \textbf{Geometry} & $\mathbf{F}$ & $\mathbf{f_1}$ & $\mathbf{f_2}$ & $\mathbf{f_3}$ & $\mathbf{s_u}$ & $\mathbf{s_u-s_l}$ & $\mathbf{S}$ $\left[\mathbf{mg^{-1}}\right]$ & $\mathbf{G_{\tau}\left(\xi_0\right)}$ \textbf{[dBi]} & $\mathbf{G_{\tau}\left(\xi_F\right)}$ \textbf{[dBi]} & $\mathbf{\Delta G_{\tau}}$ \textbf{[dB]}\\
    \hline
    \hline
         Sensitivity-opt. & $0.83$ & $0.98$ & $0.50$ & $1.00$ & $501$ & $501$ & $19.4$ & $-0.8$ & $-11.8$ & $-11.0$ \\
         Gain-optimized & $0.66$ & $0.24$ & $0.77$ & $0.15$ & $121$ & $94$ & $3.0$ & $-1.0$ & $-2.3$ & $-1.3$ \\
    \end{tabular}
\end{table*}

\begin{figure}
    \centering
    \includegraphics[width=9cm]{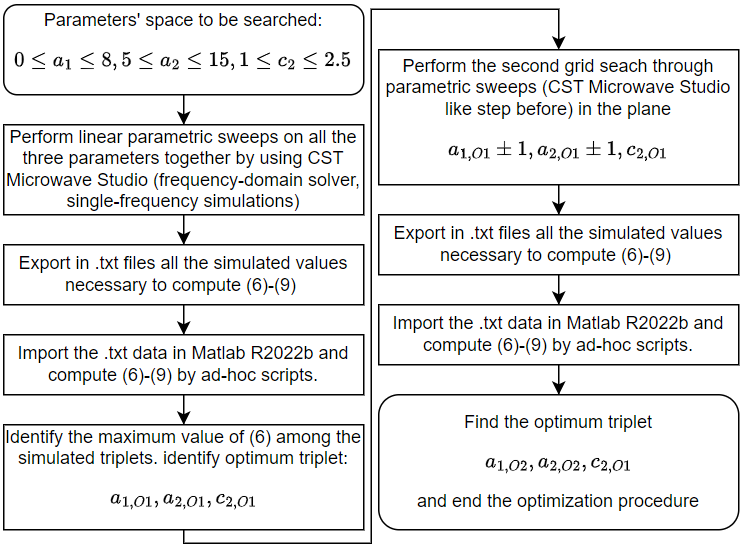}
    \caption{Flowchart detailing the two-step hierarchical grid search performed. $\left\{a_{1,O1},a_{2,O1},c_{2,O1}\right\}$ are the optimum values of the $\left\{a_{1},a_{2},c_{2}\right\}$ triplet found in the first search, and $\left\{a_{1,O2},a_{2,O2}\right\}$ are the values of $\left\{a_1,a_2\right\}$ refined in the second grid search.}
    \label{fig:designWorkflow}
\end{figure}

\begin{figure}[t]
\centering
\includegraphics[width=4.25cm]{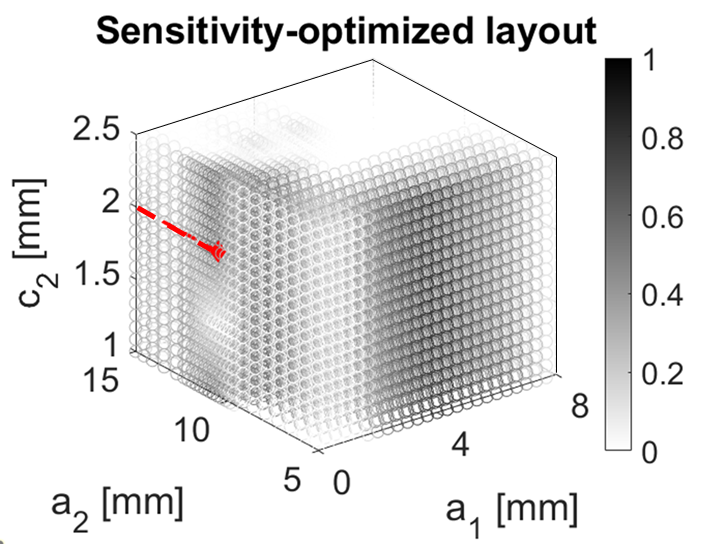}
\includegraphics[width=4.25cm]{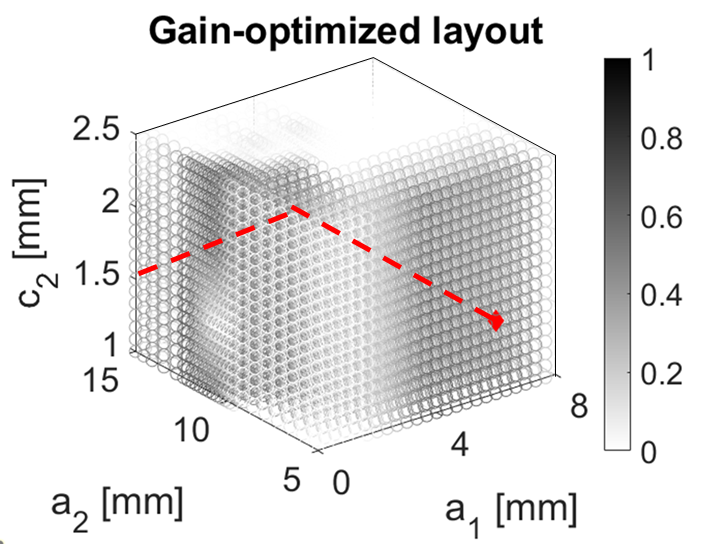}\\
\vspace{0.2cm}
\includegraphics[width=4.25cm]{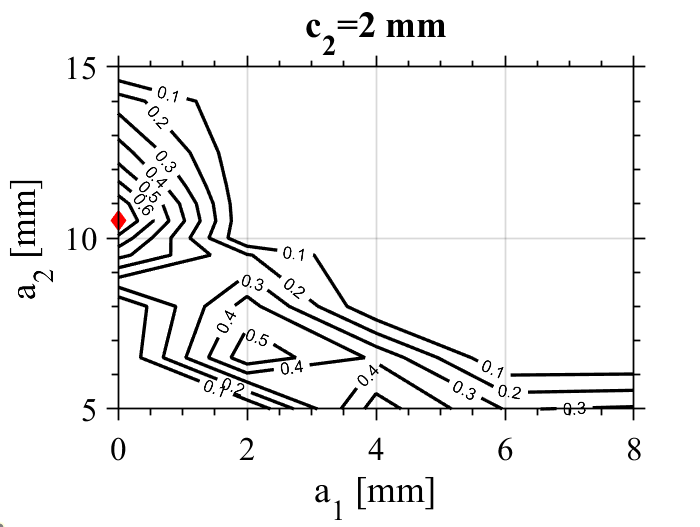}
\includegraphics[width=4.25cm]{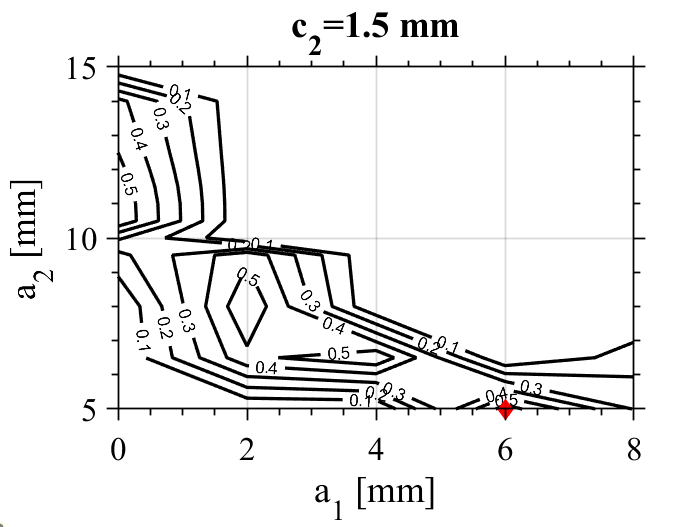}\\
\vspace{0.2cm}
\includegraphics[width=4.25cm]{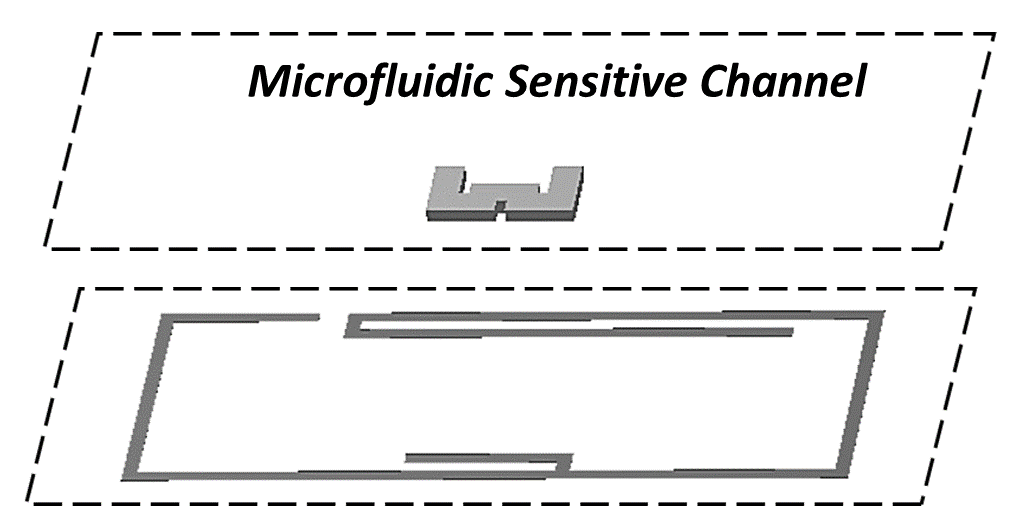}
\includegraphics[width=4.25cm]{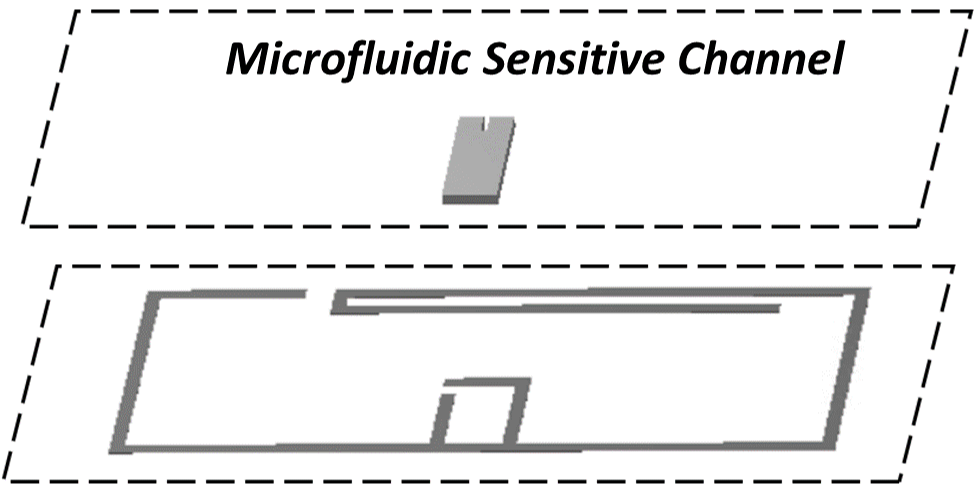}\\
(a) \qquad \qquad \qquad \qquad \qquad (b)
\caption{Numerical optimization for the two different geometries to be designed: (a) sensitivity-optimized and (b) gain-optimized. For each geometry are shown, from the top to the bottom: the parameters' space with the goal function values evaluated in the points of the hierarchical grid and the $c_2$ value of the optimal solution highlighted; the slice including the best solution; the exploded numerical model of the antenna-microfluidic system, omitting the antenna substrate and the ovoidal input and waste pads of the microfluidic.}
\label{fig:jointDesignResults}
\end{figure}

\begin{figure}
\centering
\textbf{\large{Simulated radiation gain, power transfer coefficient, realized gain and differential SC}}
\vspace{0.1cm}
\includegraphics[width=7cm]{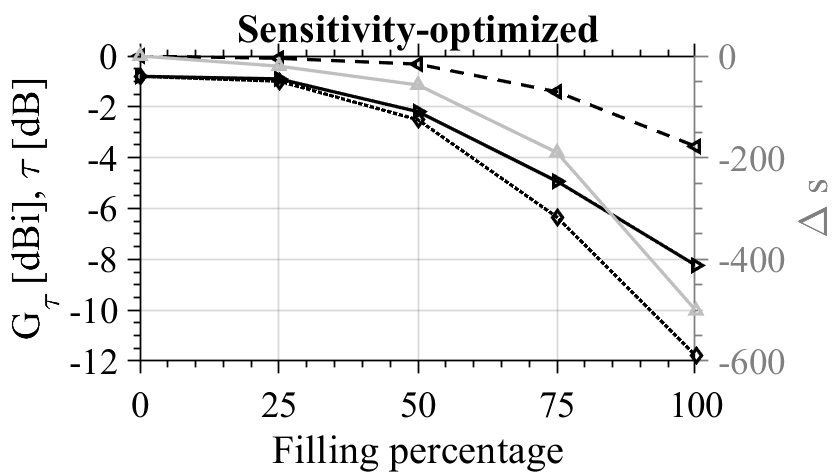}\\
\includegraphics[width=8cm]{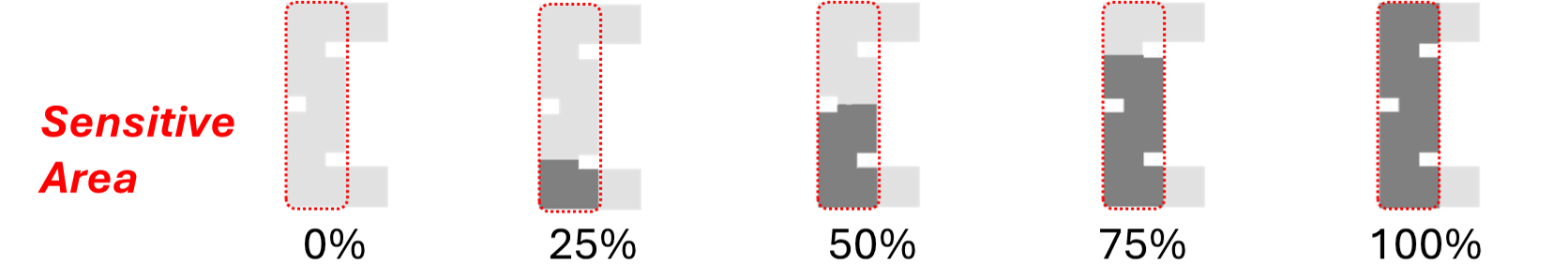}\\
\vspace{0.1cm}
\includegraphics[width=7cm]{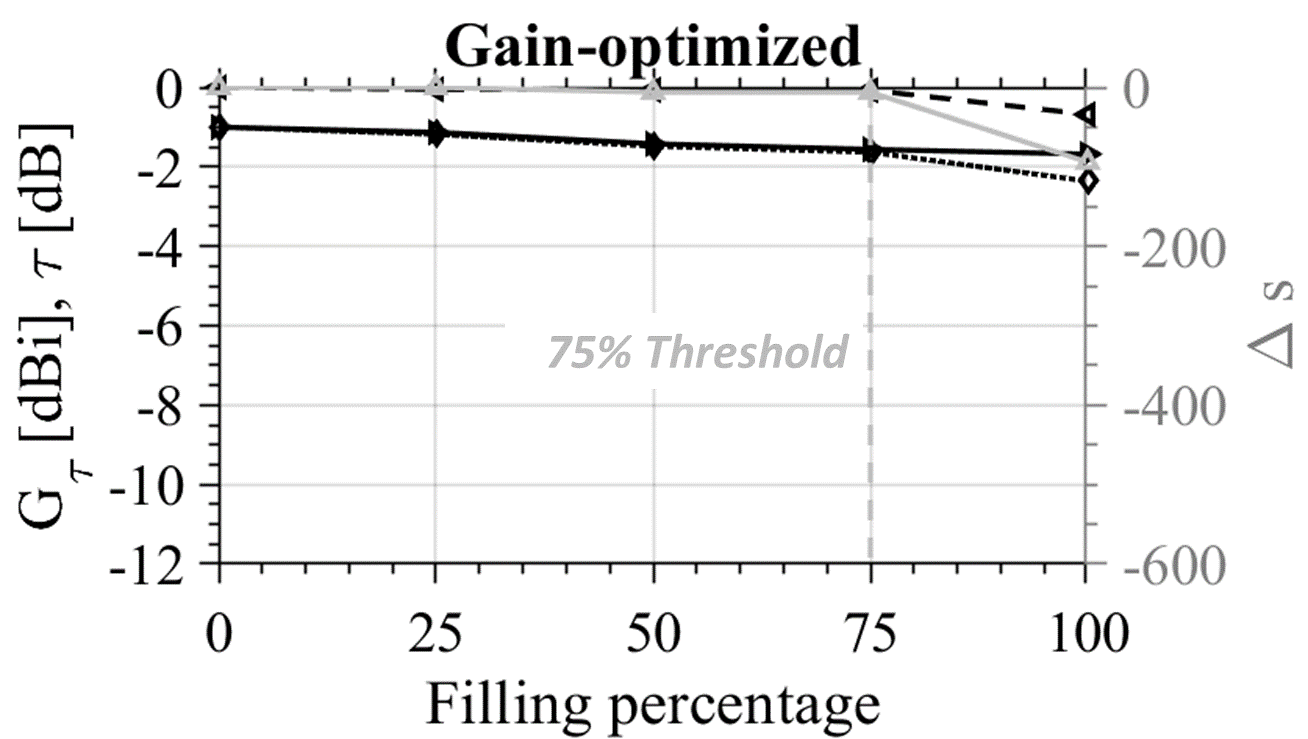}\\
\includegraphics[width=8cm]{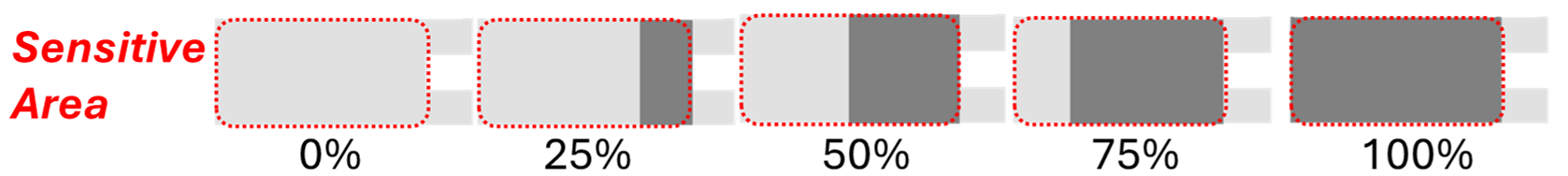}\\
\vspace{0.1cm}
\includegraphics[width=7cm]{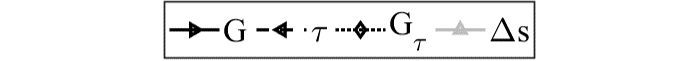}
\caption{Simulated performance of the two layouts when increasing the filling percentage of the sensitive areas. The numerical models of the loaded sensitive areas are reported in the insets and are based on experimental observation (see next Fig.~\ref{fig:redColorant}).}\label{fig:simulatedPerfo}
\end{figure}

Two geometries were jointly designed by using different sets of weights, and they are hereafter referred to as (\textit{i})~"\emph{sensitivity-optimized}" and (\textit{ii})~"\emph{gain-optimized}" solutions (Table~\ref{tab:weightsGeom}). The two different sets of weights $\left\{w_m\right\}$ were considered to optimize the sensitivity in the continuous detection of liquid (sensitivity-optimized layout) and preserve the antenna gain and, consequently, the reading distance by accepting a lower sensitivity (gain-optimized layout). As shown in the following, the second setting will be suitable as a threshold sensor.\par

The parameters' space was searched by using a hierarchical grid search, which is a largely employed technique to address nonlinear problems in low-dimensional spaces like the one under analysis \cite{Bischl23,Bianco21MultiSlope, MartinezRamon21}. The hierarchical grid can be finely refined through subsequent searches near the area of the maximum found value. For each optimization round, numerical simulations were completed by using CST Microwave Studio Suite 2023 (frequency-domain solver, single-frequency simulations), and then the results were given as input to a Matlab R2022b script to perform the hierarchical grid search. The values $\left\{G_0,S_0\right\}$ were updated at each optimization round and set to their maximum found on the grid. For the sake of completeness, the minimum values to be met by the solutions in (\ref{eq:sub2}) and (\ref{eq:sub3}) were $\left|G_{min}\right|=\left|S_{min}\right|=0$, thus making all explored geometries acceptable. Fig.~\ref{fig:designWorkflow} details the two-step hierarchical grid used to find the final antenna-microfluidic systems. In the end, the optimal geometries were further assessed through a multi-frequency simulation covering the frequency range $850-1050$~MHz to validate the joint design.\par

The numerical results of the joint design procedure for the two geometries are reported in Fig.~\ref{fig:jointDesignResults}. From Table~\ref{tab:weightsGeom}, it is possible to see that the sensitivity term in (\ref{eq:sub3}) has the highest impact on the overall value $F\left[\boldsymbol{v}\right]$. The two antenna-microfluidic geometries differ from each other and return performances in agreement with the design goals (Table~\ref{tab:simPerf}). The sensitivity-optimized geometry returns $S\sim20$~mg$^{-1}$ at the cost of a progressive reduction of the antenna gain while the serpentine is being filled with liquid. Instead, the gain-optimized geometry keeps the antenna gain near invariable (Fig.~\ref{fig:simulatedPerfo}), achieving a $G_{\tau}\left(\xi_F\right)$ value $9.5$~dB higher than the one of the sensitivity-optimized layout, whereas the sensor-code changes only when the filling percentage of the sensitive channel exceeds $75\%$ of the maximum value and, hence, the device can be considered as a threshold sensor capable of discriminating between the unloaded and nearly-fully-loaded microfluidic.

\section{Prototypation and Measurements}\label{sec:prototypation}

\begin{figure}[!t]
\centering
\includegraphics[height=4cm]{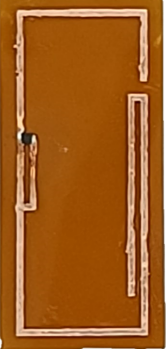}
\qquad
\includegraphics[height=4cm]{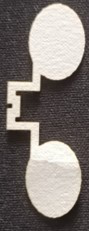}
\qquad
\includegraphics[height=4cm]{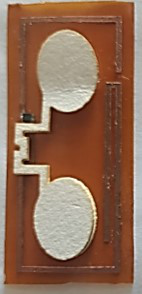}\\
(a) \qquad \qquad \qquad (b) \qquad \qquad \qquad (c)\\
\vspace{0.1cm}
\includegraphics[height=4cm]{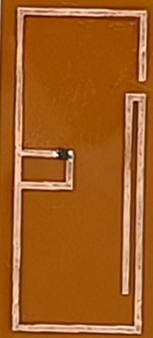}
\qquad
\includegraphics[height=4cm]{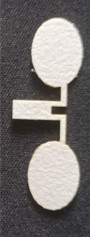}
\qquad
\includegraphics[height=4cm]{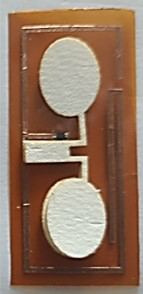}\\
(d) \qquad \qquad \qquad (e) \qquad \qquad \qquad (f)\\
\caption{Prototypes of the two antenna-microfluidic systems. Sensitivity-optimized layout: (a) RFID antenna, (b) microfluidic channel, and (c) whole system. Gain-optimized layout: (d) RFID antenna, (e) microfluidic channel, and (f) whole system.}
\label{fig:prototyopes}
\end{figure}

\begin{figure}[tp]
\centering
\includegraphics[width=8.5cm]{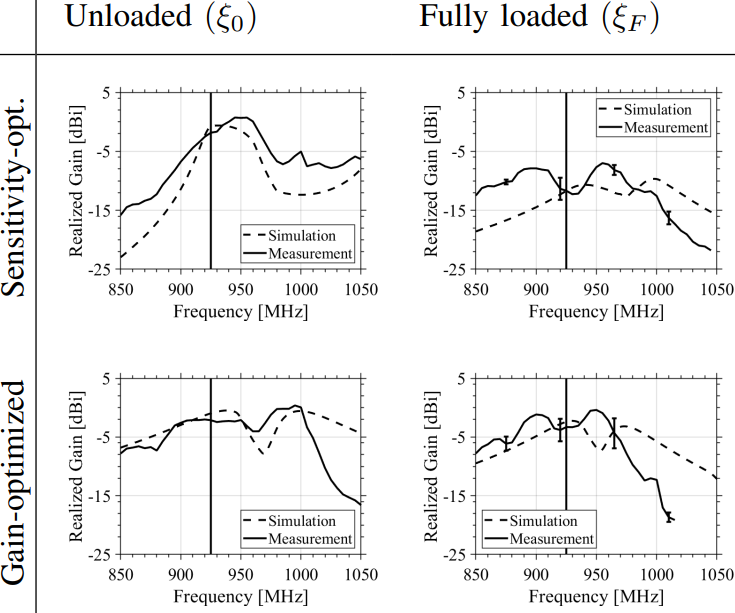}
\caption{Electromagnetic measurements of the antenna-microfluidic system compared with the simulated ones (average over $5$~measurements). The working frequency of $925$~MHz is highlighted by a vertical line and the standard deviation is not reported if it is lower than $0.5$~dB.}
\label{fig:EMMeasures}
\end{figure}

The two prototypes were fabricated by using a $35$-$\mu$m-thick copper layer attached to a Kapton film as in the simulations and, then, cut by a milling machine ($4$MILL$300$ATC by Mipec~\cite{Mipec23}). Based on the authors' prior experience with filter papers~\cite{Fiore23,Riente23} and the preliminary investigation we reported in \cite{Bianco24EUCAP}, the microfluidic was manufactured by manually superposing two identical layers of Whatman CF$4$ filter paper (from \cite{Ghiaroni23}; single-layer thickness at $53$~KPa: $482$~$\mu$m: water absorption: $49.9$~mg/cm\textsuperscript{2}). \par

The filter paper was cut through a $40$W USB CO\textsubscript{$2$} Laser Engraving Machine Cutter (by Vevor~\cite{Vevor23}) having a position accuracy of $10$~$\mu$m and engraving precision of $2500$~dpi. The designed microfluidic channel was connected to two identical and ellipsoidal input and waste pads. The microfluidic was fixed to the Kapton substrate by using a thin polypropylene tape without contacting the conductor trace in any point since similar arrangements are known not to alter the self-tuning sensing~\cite{Bianco20Near}. For the measurements, four antenna prototypes (two per geometry) and twenty-two layers of filter paper (i.e., eleven double-layer microfluidic channels per geometry) were employed. Each paper layer was cut by the laser cutter, and, by superimposing two layers, one microfluidic prototype was obtained; then, the double-layer microfluidic channels were utilized for single-use, whereas the antennas were not affected from usage and could be used again after replacing the microfluidic channel and some minutes of dry time. The antenna-microfluidic prototypes are shown in Fig.~\ref{fig:prototyopes}.\par 

As the first test, the realized gain was measured through a Tagformance station (by Voyantic; linearly-polarized interrogation antenna AN-FF-WB; $55$~cm of interrogation distance) in the extreme conditions of unloaded and loaded tags. Good agreement and stability were observed between measurements and simulations in both conditions (Fig.~\ref{fig:EMMeasures}), allowing for a theoretical reading range $>1$~m when interrogating the tags with a circularly-polarized antenna and $1$~W EIRP. Notably, the gain-optimized geometry achieves gain degradation of just $\Delta G_{\tau}\simeq-1$~dB between the two boundary conditions in contrast with a degradation $\Delta G_{\tau}\simeq-10$~dB in the case of the sensitivity-optimized system layout.\par

\begin{figure}[tp]
\centering
\includegraphics[width=7cm]{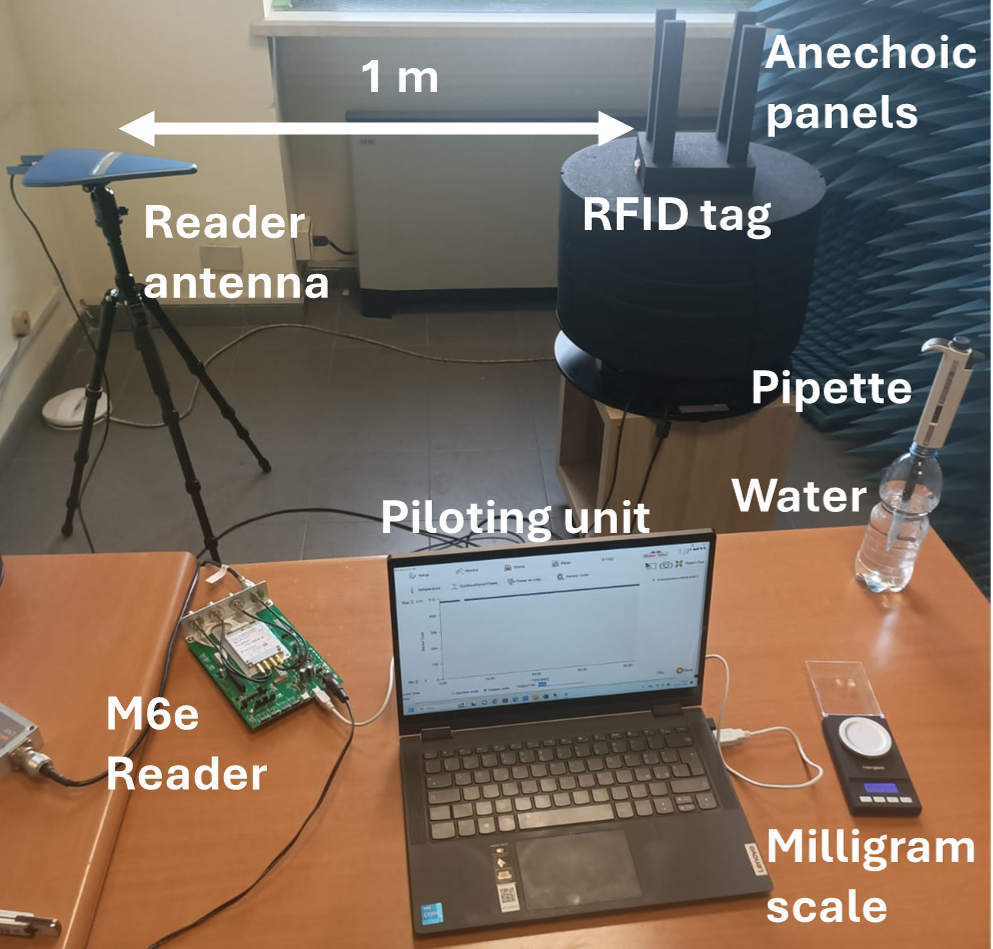}
\caption{Experimental set-up for the sensor code measurements with the highlighted reader-tag distance of $1$~m.}
\label{fig:scMeasureSetup}
\end{figure}

\begin{figure}[tp]
\centering
\begin{tabular}{ccccc}
\includegraphics[height=3.5cm,width=1.3cm]
{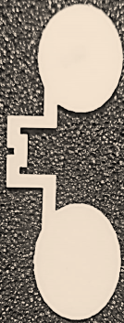}&\includegraphics[height=3.5cm,width=1.3cm]{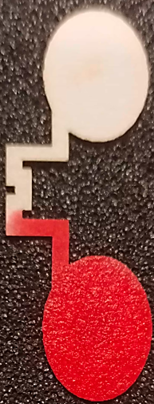}&\includegraphics[height=3.5cm,width=1.3cm]{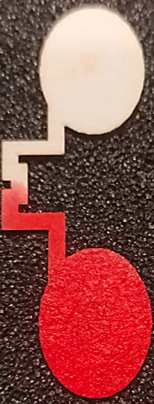}&\includegraphics[height=3.5cm,width=1.3cm]{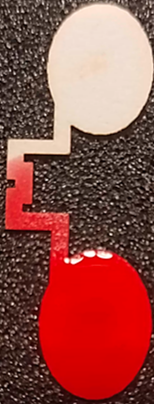}&\includegraphics[height=3.5cm,width=1.3cm]{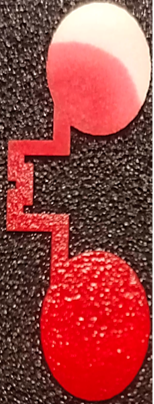}\\
$0\%$ & $25\%$ & $50\%$ & $75\%$ & $100\%$ \\
\includegraphics[height=3.5cm]
{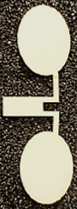}&\includegraphics[height=3.5cm,width=1.3cm]{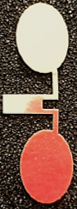}&\includegraphics[height=3.5cm,width=1.3cm]{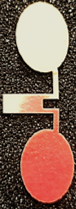}&\includegraphics[height=3.5cm,width=1.3cm]{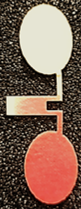}&\includegraphics[height=3.5cm,width=1.3cm]{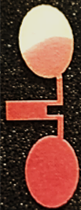}
\end{tabular}
\caption{Visual examples showing of the progression of coloured water into the microfluidics sensitive channel (filling percentage) of the sensitivity-optimized (above) and the gain-optimized (below) layouts.}
\label{fig:redColorant}
\end{figure}

\begin{figure}[t]
\centering
\includegraphics[width=8cm]{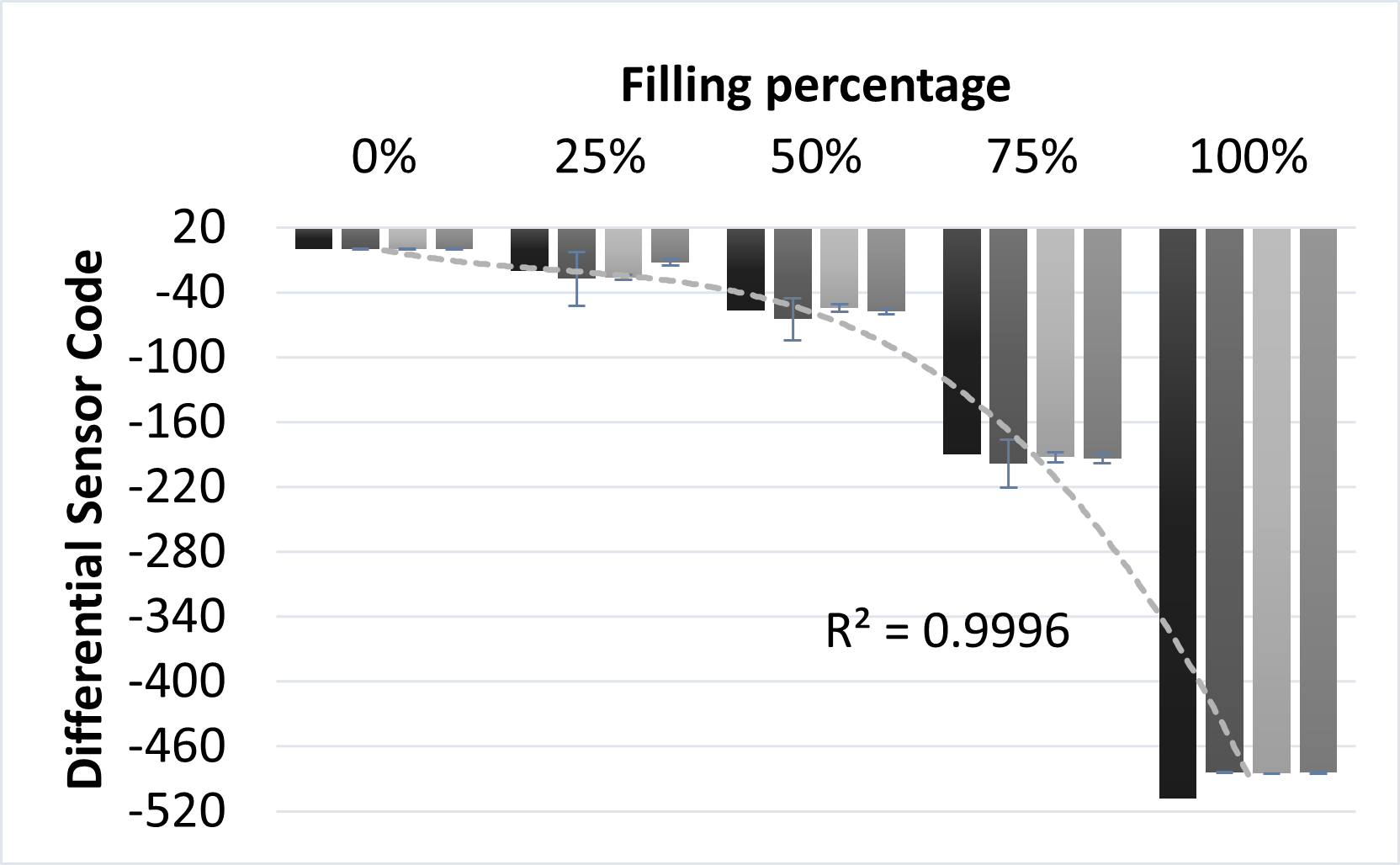}\\
(a)\\
\includegraphics[width=8cm]{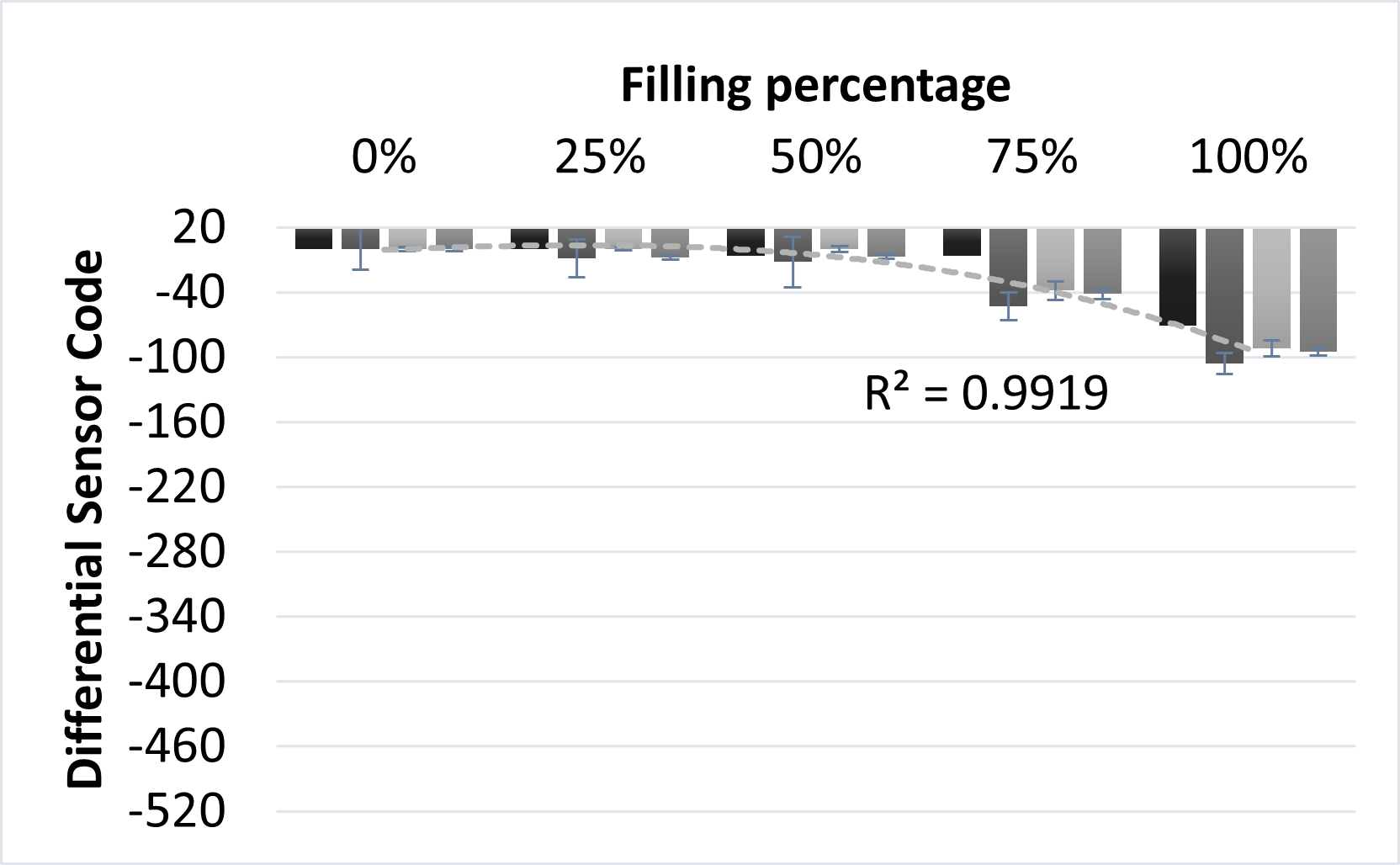}\\
(b)\\
\includegraphics[width=8.5cm]{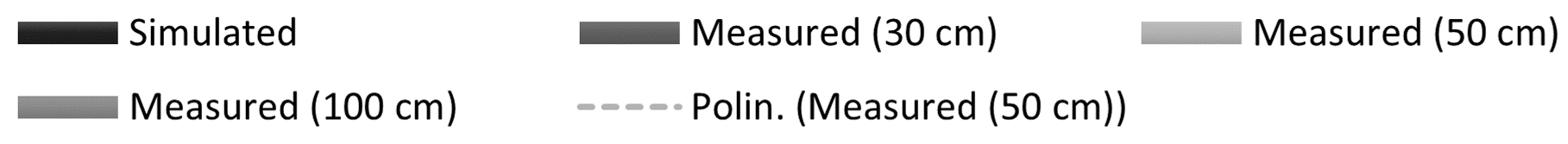}\\
\caption{Comparison between the simulated and measured $\Delta s$ values of the two geometries: (a) sensitivity-optimized and (b) gain-optimized. Cubic polynomial trend lines fitting the data read from the intermediate distance ($50$~cm) are reported with the corresponding R$^2$ values.}
\label{fig:scMeasuresComp}
\end{figure}

\begin{table*}[t]
\caption{Measured (meas.) performance of the prototypes and absolute value of the difference (diff.) with the simulated values. The measurements are averaged on the three reading distances.}\label{tab:measSimComp}
    \centering
    \begin{tabular}{l||r|r|r|r|r|r|r|r|r|r|r|r}
         & \multicolumn{2}{c|}{$\mathbf{s_u}$} & \multicolumn{2}{c|}{$\mathbf{s_u-s_l}$} & \multicolumn{2}{c|}{$\mathbf{S}$ $\left[\mathbf{mg^{-1}}\right]$} & \multicolumn{2}{c|}{$\mathbf{G_{\tau}\left(\xi_0\right)}$} & \multicolumn{2}{c|}{$\mathbf{G_{\tau}\left(\xi_F\right)}$} & \multicolumn{2}{c}{$\mathbf{\Delta G_{\tau}}$}\\
    \hline
     \textbf{Geometry} & Meas. & Diff. & Meas. & Diff. & Meas. & Diff. & Meas. & Diff. & Meas. & Diff. & Meas. & Diff. \\
    \hline
    \hline
         Sensitivity-opt. & $489\pm 1$ & $12$ & $484\pm 27$ & $17$ & $18.8$ & $0.6$ & $-1.8$ dBi & $1.0$ dB & $-11.7$ dBi & $0.1$ dB & $-9.9$ dB & $1.1$ dB \\
         Gain-optimized & $307\pm 13$ & $186$ & $98\pm 18$ & $4$ & $3.1$ & $0.1$ & $-2.1$ dBi & $1.1$ dB & $-3.3$ dBi & $1.0$ dB & $-1.2$ dB & $0.1$ dB \\
    \end{tabular}
\end{table*}

Afterwards, the sensing capability of the prototypes was assessed through a M6e (by ThingMagic; $30$~dBm of interrogation power), the same AN-FF-WB antenna from the measurements above, and custom software (from previous works~\cite{Bianco20Near}) for the retrieval of the SC. Liquid quantity was increased by using a micropipette (M1000 by QWork) and the weight was controlled by using a digital scale (nominal resolution: $1$~mg) to check the system's weight before measurements (Fig.~\ref{fig:scMeasureSetup}). A maximum error on liquid weight of $5\%$ was accepted due to the imperfect manual pipetting. Three measurement distances ($30$, $50$, and $100$~cm) were tested to investigate eventual effects of the interrogation distance on the digital metric. $10$~SC samples were averaged for the measurement-simulation comparison based on the convergence error analysis in~\cite{Naccarata22}. The differential SC from Sec.~\ref{sec:backgroundSelftuning} was used to drop out IC-dependent baselines. The simulations-measurements comparison was referred to the filling percentage of the microchannels, measured by weighting (stable $s_l$ achieved, adding more water did not cause any SC variation; $1050$~mg for the sensitivity-optimized layout and $900$~mg for the gain-optimized layout; weights measured after the water reached the sensitive channel). Red food colorant was employed~\cite{Riente23} in order to visually emphasize the water quantity present in the microfluidic. Fig.~\ref{fig:redColorant} shows the fluid progressing similarly to the EM numerical models in Fig.~\ref{fig:simulatedPerfo}.\par

Despite the very different SC ranges used by the two geometries, good measurement-simulation agreement is always observed~(Table~\ref{tab:measSimComp}). Fig.~\ref{fig:scMeasuresComp} depicts the comparison with the data measured from the three distances, showing good robustness of the $\Delta s$ metric when changing the interrogation distance. The trend of the SC when increasing the water follows a cubic trend line (R$^2>0.99$) although the EM simulations did not capture the phenomenon for the gain-optimized geometry due to the simplyfied numerical model of the antenna-microfluidic system. As expected, the sensitivity-optimized geometry exploited the largest sensor-code range, achieving the best overall sensitivity ($S=18.8$ units of sensor code per water milligram) at the expense of the radiation gain. The sensor code response of the "gain-optimized" geometry stays flat until the filling percentage of the serpentine reaches~$75\%$, in good agreement with the simulation, with a variation of $\Delta s\simeq100$ between the two extremal loading conditions.\par

\section{Conclusion}\label{sec:conclusion}

In this work, the joint design of an antenna and a microfluidic channel has been proposed, formulated, and experimentally validated. When comparing the design examples with the systems for wireless liquid sensing in Table~\ref{tab:biblio}, our examples monitored a larger quantity of fluid with the added benefits of relying on a dedicated, digitalized metric~\cite{Occhiuzzi16}; on the other hand, the main con of the proposed approach consists in the complexity of the EM modeling of the microfluidics since inadequate modeling can result in design failure. In the joint design, the balance between sensitivity and communication performance can be easily modulated by properly selecting the optimization weights, thus offering the flexibility to adapt the system to the specific application of interest. \par

The design method proposed in this article can be used for optimizing UHF RFID antenna-microfluidic systems to address open issues in healthcare, like point-of-care devices for monitoring hyperidrosis~\cite{Walling11} or sweating during domestic exercise~\cite{Bianco23PoC}, and food sectors, like devices estimating food quality through "drip loss" quantification~\cite{Mostaccio23Food,Manheem23}, eventually by using ad-hoc RFID totems~\cite{Occhiuzzi20}. Yet, the microfluidic model we employed is simple and not representative of more complex microfluidic circuits or fast flows of liquids; in this second case, multiphysics simulators could be employed, to exploit the joint design when the simple modeling described in Sec.~\ref{sec:emmodeling} does not hold, for instance, to account quantitatively for a significant wicking effect of the paper. Moreover, the choice of the filter paper was not investigated throroughly and was performed based on prior authors' experience to complete the design example. Furthermore, the quantity of the liquid to be monitored is only governed by the microfluidic channel through the assumption that the input and waste pads and the microfluidic path can carry the quantity of interest in the sensitive area. Future research directions will address those open issues and explore the deployment of jointly designed antenna-microfluidic systems to exploit and refine the joint design technique.\par

\section*{Acknowledgements}\label{sec:acknowledge}
The authors thank Prof. Fabiana Arduini, Dr Vincenzo Mazzaracchio and Dr Luca Fiore (all affiliated with the Department of Chemical Science and Technologies, Tor Vergata University of Rome) for their fundamental collaboration in fabricating the microfluidic channels.

\bibliographystyle{IEEEtran}
\bibliography{main} 

\end{document}